\documentstyle[osa,manuscript]{revtex}

\begin{document}


\begin{center}
{\bf Statistical Interpretation of Joint Multiplicity Distributions
of Neutrons and Charged-particles}
\vskip.25in
{J. T\~oke, D.K. Agnihotri, W. Skulski, and W.U. Schr\"oder\\}
{\it Department of Chemistry and Nuclear Science Research
Laboratory,\\
University of Rochester, Rochester, New York 14627}
\end{center}
\bigskip
\begin{abstract}

Experimental joint multiplicity distributions of neutrons and charged
particles provide a striking signal of the characteristic decay
processes of nuclear systems following energetic nuclear reactions.
They present, therefore, a valuable tool for testing theoretical models for
such decay processes. The power of this experimental tool is
demonstrated by a comparison of an experimental joint multiplicity
distribution to the predictions of different theoretical models
of statistical decay of excited nuclear systems. It is shown that, while
generally phase-space based models offer a quantitative description
of the observed correlation pattern of such an
experimental multiplicity distribution, some
models of nuclear multifragmentation fail to account for
salient features of the observed correlation.  

\end{abstract}

\pacs{PACS numbers: 25.70.-z,25.70.Lm,25.70.Pq}
\newpage


Nuclear
multifragmentation\cite{Gel87,Bor90,Mor93,Ber88,Fri88,Aic91,smm,mmmc},
the production of multiple intermediate-mass fragments (IMFs) in
individual  reaction events, has been one of the central issues in
intermediate-energy heavy-ion  reactions in the last decade. The
theoretical effort \cite{Ber88,Fri88,Aic91,smm,mmmc} to understand
this phenomenon is largely driven by an expectation that this process
may reveal a macroscopic behavior of nuclear matter at very high
excitation energies that is qualitatively different from its behavior
at lower excitation energies. Of particular interest in this context
is the prospect of probing the nuclear liquid-gas phase transition,
which has often been associated with multiple IMF production. Parallel
with purely theoretical effort, a search has been conducted for
experimental signatures that could be connected to certain IMF
production scenarios by means of simple and reliable simulation
calculations.\cite{tok1,mor1,mor2} At the present stage of this
research, select partial sets of experimental observations are
apparently consistent with different reaction scenarios and mutually
exclusive physical concepts. For further progress in the understanding
of the reaction mechanism, it is therefore essential to identify
reliable experimental observations that would challenge some but not
other models and propositions. It now appears, that the directly
measured joint distribution of neutron and charged-particle
multiplicities is one such reliable observable. The significance of
this observable, in general, and for the ongoing discussion regarding
the character of 
multi-fragmentation\cite{tok1,mor1,mor2,botv1,botv2,tsang,wojtfld}, in
particular, has gone so far largely unnoticed, due to a seemingly
trivial character of the information contained in the above joint
distributions. The present work demonstrates the discriminative power
of the combined multiplicity observable in two sample analyses.

Fig.~1 shows a typical joint distribution of neutron
and light charged-particle multiplicities, $m_n$ and $m_{LCP}$, as
observed in the $^{209}$Bi+$^{136}$Xe  reaction at E/A=28 MeV and
reported on earlier.\cite{WUS92,neck} The notable feature of the
observed distribution is the presence of a well-defined correlation
ridge with a characteristic bend around ($m_n$,$m_{LCP}$) $\approx$
(15,0). When following the crest of the correlation ridge in
Fig.~1 beginning from the origin of the plot, one first
encounters a rather long segment running straight and parallel  to the
$m_n$ axis. Along this first segment, $m_{LCP}\approx 0$ and, hence,
the role of the charged-particle emission is insignificant. The latter
decay mode becomes a factor only when the measured neutron
multiplicity exceeds rather significant values of the order of
$m_n$=15 (corresponding to a true, efficiency-corrected value of
$m_n\approx$30). At above multiplicities of $m_n\approx$ 15, the crest
of the correlation ridge is seen to part with the neutron multiplicity
axis and run at a constant angle with respect to the $m_n$ axis, until
a saturation in both, neutron and LCP multiplicities is reached. It is
the presence of the first, $m_{LCP}\approx 0$ segment of the joint
distribution and its appreciable length that poses a challenge to some
propositions regarding IMF production.

\vspace*{13cm}
\includegraphics{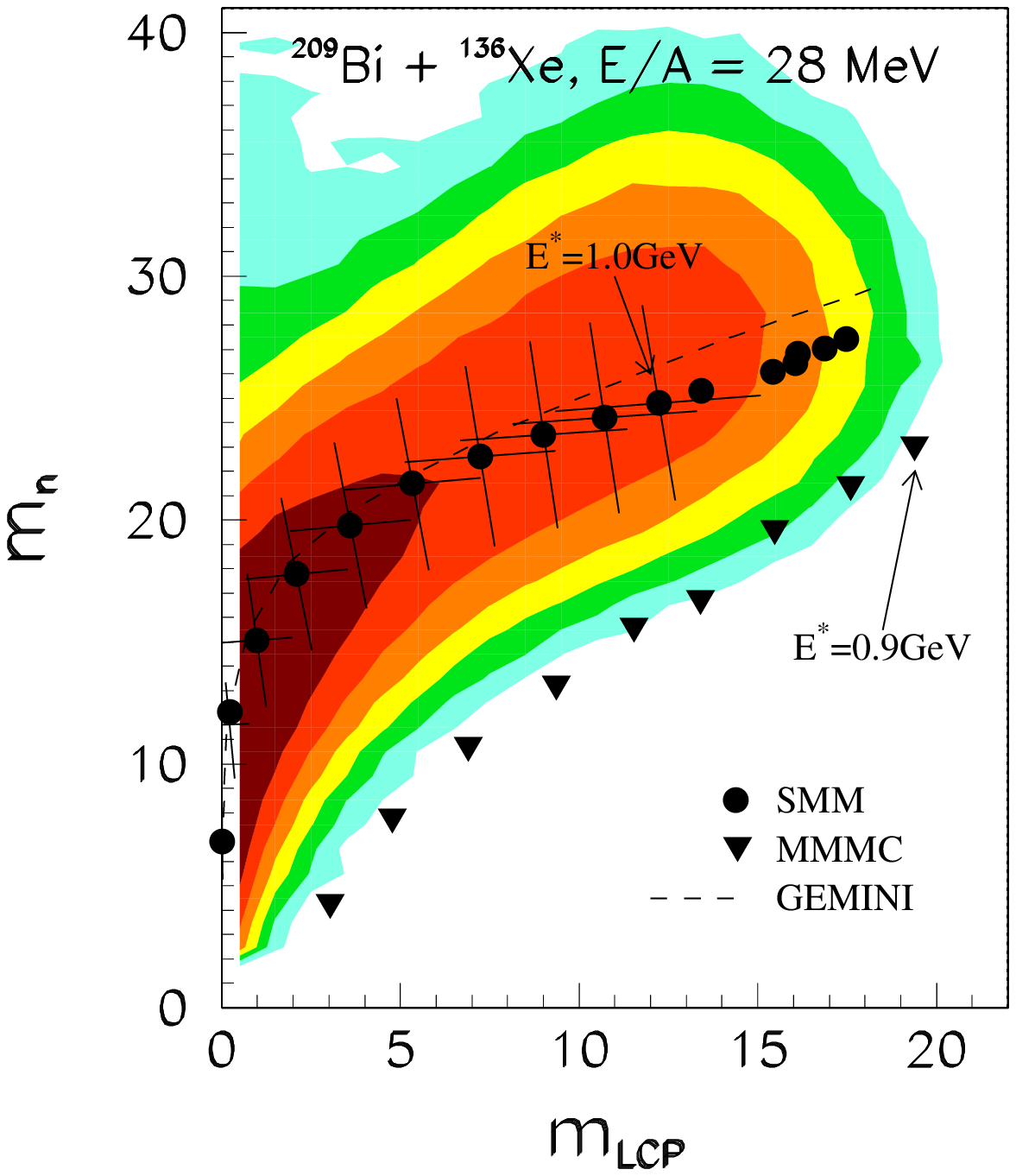}
{\footnotesize
\begin{quotation}
\noindent
Fig. 1. Experimental$^9$ logarithmic (base 2) contour plot of the joint
distribution of
neutron and charged-particle multiplicities. Trends
predicted by the codes SMM,$^7$ MMMC,$^8$, and GEMINI$^{20}$
are illustrated by circles, squares, and a dashed line,
respectively. The slanted bars attached to some of the theoretical
data points illustrate the (efficiency-corrected) 
local orientation and lengths (FWHM) of the
major and minor axes of the $m_n$-$m_{LCP}$ covariance tensor
for a fixed total excitation energy.
\end{quotation}}
\bigskip

It is important to emphasize
that the above two-segment topography of the correlation ridge in the
joint $m_n$ {\it vs.} $m_{LCP}$ distribution has been observed for
many systems and at different bombarding energies. The pertinent
feature discussed above is also consistent with earlier
observations\cite{Jia89} of average neutron and charged-particle
multiplicity correlations for other systems. As a matter of fact, the
same characteristic shape has been observed in any experiment where
such joint distributions or average correlations were measured. These
experiments include those studying reactions induced by relativistic
protons and antiprotons.\cite{benoit}

The behavior seen in Fig.~1 is consistent with a
dominantly statistical, phase-space governed, emission scenario of
neutrons and light charged products. One expects that the lack of a
Coulomb barrier for neutrons would lead to a strong dominance of the
neutron emission channels at low excitation energies and, hence, to
the presence of the first, $m_{LCP}\approx 0$ segment of the
correlation ridge. Simulation calculations confirm\cite{WUS92} that
phase-space models, such as the evaporation model\cite{gemini} GEMINI
and the Copenhagen model\cite{smm} of simultaneous multi-fragmentation
(SMM), reproduce quite accurately the topography of the correlation
ridge in Fig.~1 and, especially, the location of the 
crest line of the experimental ridge. As an example, the results of
SMM\cite{smm} and  GEMINI\cite{gemini} calculations are indicated in
Fig.~1 by solid dots and the dashed line, respectively.  
The ``error'' bars attached to the dots illustrate the orientation and
the lengths (in terms of FWHM) of the main axes of the theoretical
correlation tensor of the neutron and LCP multiplicities for fixed
excitation energies. In the above model calculations, it was assumed
that neutrons and light-charged particles are emitted 
from fully-accelerated projectile- and target-like fragments produced
in the primary dissipative collision. Such an assumption is justified
by the experimental observation\cite{Que93,benoit} that, 
for heavy-ion reactions of interest here
and low kinetic-energy losses, particle emission patterns are
characteristic of dominantly statistical emission from two
equilibrated sources, with only a weak contribution of pre-equilibrium
emission to the particle yield. Subsequently, 
the raw
theoretical predictions have been corrected for the efficiency of the
Rochester RedBall neutron multiplicity meter,\cite{baldwin} used in
the measurement of the joint multiplicity distribution shown in
Fig.~1.  These efficiencies were calculated using a modified version
of the Monte Carlo code DENIS,\cite{denis} which has been calibrated 
using various sets of experimental data.\cite{baldwin} For the
assumed binary kinematics, they are in the range of 55\% -- 61\%.
   
In view of the above, an interpretation of the correlation ridge seen 
in Fig.~1 in terms of a dominantly statistical process
is well warranted. This is especially true for low total
excitation energies, on which the conclusions of the
present paper rely.

The importance of the experimental measurements of joint distributions
of neutron and LCP multiplicities is demonstrated by the fact that not
all prominent models for the decay of hot nuclear matter account  for
the gross trends observed in the experimental data. Quite obviously
then and, possibly, contrary to one's intuition, these  trends are not
trivial at all. Notably, the Berlin microcanonical fragmentation
model, as implemented in the Metropolis Monte Carlo code\cite{mmmc}
MMMC, does not reproduce satisfactorily the first, relatively long
segment of the $m_n$ {\it vs.} $m_{LCP}$ correlation ridge parallel to
the $m_n$ axis. Instead, as illustrated by the triangles in
Fig.~1, the model predicts a correlation ridge that is
characterized by an approximately straight proportionality between
average multiplicities of neutrons and charged-particles. We note,
that calculations for all three models were performed for identical
initial conditions, to provide for a meaningful comparison.  

It is clear from Fig.~1 that MMMC calculations largely
underestimate the role of neutron emission channels in the decay of
excited nuclear systems and apportion the thermal energy incorrectly to
the different particle types. While the physics captured by the MMMC,
on the one hand, and the SMM, on the other hand, appears to be very 
similar, these two codes differ strongly in the way they treat 
fragment excitation and neutron
emission. In the SMM, neutrons are allowed to be emitted sequentially
from thermally equilibrated
excited primary fragments, whereas the MMMC requires {\it all}
the emitted neutrons to be in a narrow band of continuum states
simultaneously at one particular moment in time, while
the core fragments are essentially cold. Obviously, the latter process 
is statistically 
much less likely than sequential emission
from hot fragments. It is this difference in the
treatment of neutron emission that makes the predictions by the MMMC
for the $m_n$ {\it vs.} $m_{LCP}$ correlations qualitatively different
from the predictions by the SMM for the same experimental observable.   
Furthermore, in the MMMC, the neutron multiplicity is largely
determined by the choice
of weight factors describing different multiplicities of
``evaporated'' neutrons. Conceptually, these
weight factors are to represent the density of micro states of
the collection of all ``evaporated'' neutrons at a given total energy
of this collection. However, in the MMMC, they are  
approximated by

\begin{equation}
\label{eq:weight}
W_p\propto V_p^{m_{ev}},
\end{equation}

\noindent where $V_p$ is the volume of a spherical shell in momentum
space, enclosed between $p_{min} = \sqrt{2U_n}$ ($U_n$ = 50 MeV being
the depth of the single-neutron potential well) and a ``suitably
chosen''\cite{mmmc} $p_{max}$. In Eq.~\ref{eq:weight}, $m_{ev}$ is the
multiplicity of ``evaporated'' neutrons.  The momentum distribution
described by Eq.~\ref{eq:weight}, corresponds not to a definite energy
but to a broad spectrum of the total energy of all ``evaporated''
neutrons and, hence, $W_p$ are not microcanonical, as are most (but
not all remaining) weight factors in the MMMC. One notes, that in the
MMMC, the competition by channels with various neutron multiplicities
is largely decided by what is termed\cite{mmmc} a ``suitable choice''
of a critical parameter $p_{max}$ in a particular (non-microcanonical)
parameterization of $W_p$ via Eq.~\ref{eq:weight}. Therefore, the
discrepancy between the MMMC prediction, on the one hand, and the
predictions by the codes SMM and GEMINI, on the other hand, cannot be
traced to any physical effects, but rather to the accuracy of
mathematical/physical shortcuts adopted by the MMMC. 
  
The utility of the joint multiplicity distributions ($m_n$ {\it vs.} 
$m_{LCP}$) as a powerful tool for scrutinizing theoretical concepts is
further demonstrated in the context of pseudo-Arrhenius plots for
multi-IMF distributions. In an ongoing discussion, the merits of a
proposed statistical interpretation of pseudo-Arrhenius plots for
nuclear multi-fragmentation have been
debated.\cite{tok1,mor1,mor2,botv1,botv2,tsang,wojtfld,newmor}. 
Two types of such plots have been considered in the published literature.  
In the first one,\cite{mor1} the logarithm of an inverse ``binomial
probability'' ln($1/p$) is plotted versus the inverse square root of
the transverse kinetic energy $1/\sqrt{E_t}$ of charged reaction
products. In the second realization,\cite{mor2} the average IMF
multiplicity $<m_{IMF}>$ is plotted versus $1/\sqrt{E_t}$. The
``binomial probability'' $p$ has been introduced\cite{mor1} based on
an experimental discovery that IMF multiplicity distributions for
any given transverse energy are well approximated by binomial
distributions:

\begin{equation}
\qquad P_n^m(p) = {m!\over n!(m-n)!}p^n(1-p)^{m-n}\;,
\label{eq_bindist}
\end{equation}

\noindent where $m$ and $p$ are the number of trials and the
probability for success in any one of these trials, respectively. The
transverse energy is defined as $E_t=\Sigma E_ksin^{2}(\Theta_k)$,
where the summation extends over all charged products, excluding
fission fragments and projectile- or target-like residues, and
$\Theta_k$ is the emission angle of the $k$-th product. The declared,
rationale\cite{mor1,mor2} for these kinds of plots is the expectation
that  for statistical IMF emission, the emission probability should be
proportional to the Boltzmann factor $e^{-B/T}$, where $B$ is an
effective emission barrier and $T$ is the nuclear temperature. This
rationale relies critically on the validity of the assumption that the
nuclear temperature $T$ is proportional to $\sqrt{E_t}$, which
presumes a direct proportionality between $E_t$ and the thermal
excitation energy $E^*$. These assumptions also entail that $E^*$ is
proportional to the square of the temperature, as is proper for a
Fermi gas. The validity of the two above assumptions determines the
appropriateness of an interpretation of pseudo-Arrhenius plots in
terms of thermal scaling, proposed in a series of recent
papers\cite{mor1,mor2,botv1,botv2,newmor}. For example, had $E_t$
depended quadratically or exponentially on $E^*$, there would have
been no obvious justification for choosing  an abscissa variable
$1/\sqrt{E_t}$ for the pseudo-Arrhenius plots. Accordingly, these
papers rely on limited simulation calculations, which always
explicitly state and take for granted, a strict proportionality
between $E_t$ and $E^*$. However, the character of the joint
multiplicity distributions discussed above demonstrates unambiguously
the lack of even an approximate, or average, proportionality between
$E_t$ and $E^*$. This is so, because neutrons are not included in the
experimental definition of the transverse energy $E_t$ in the quoted
work, whereas neutron emission is seen and understood to be
essentially the only significant decay channel available to
medium-weight or heavy nuclear systems at excitation energies of up to
several hundred MeV. As a result, $E_t\approx 0$ for the first few
hundreds of MeV of the total excitation energy, precluding direct
proportionality between $E_t$ and $E^*$ at higher excitation energies.

\vspace*{10cm}
\includegraphics{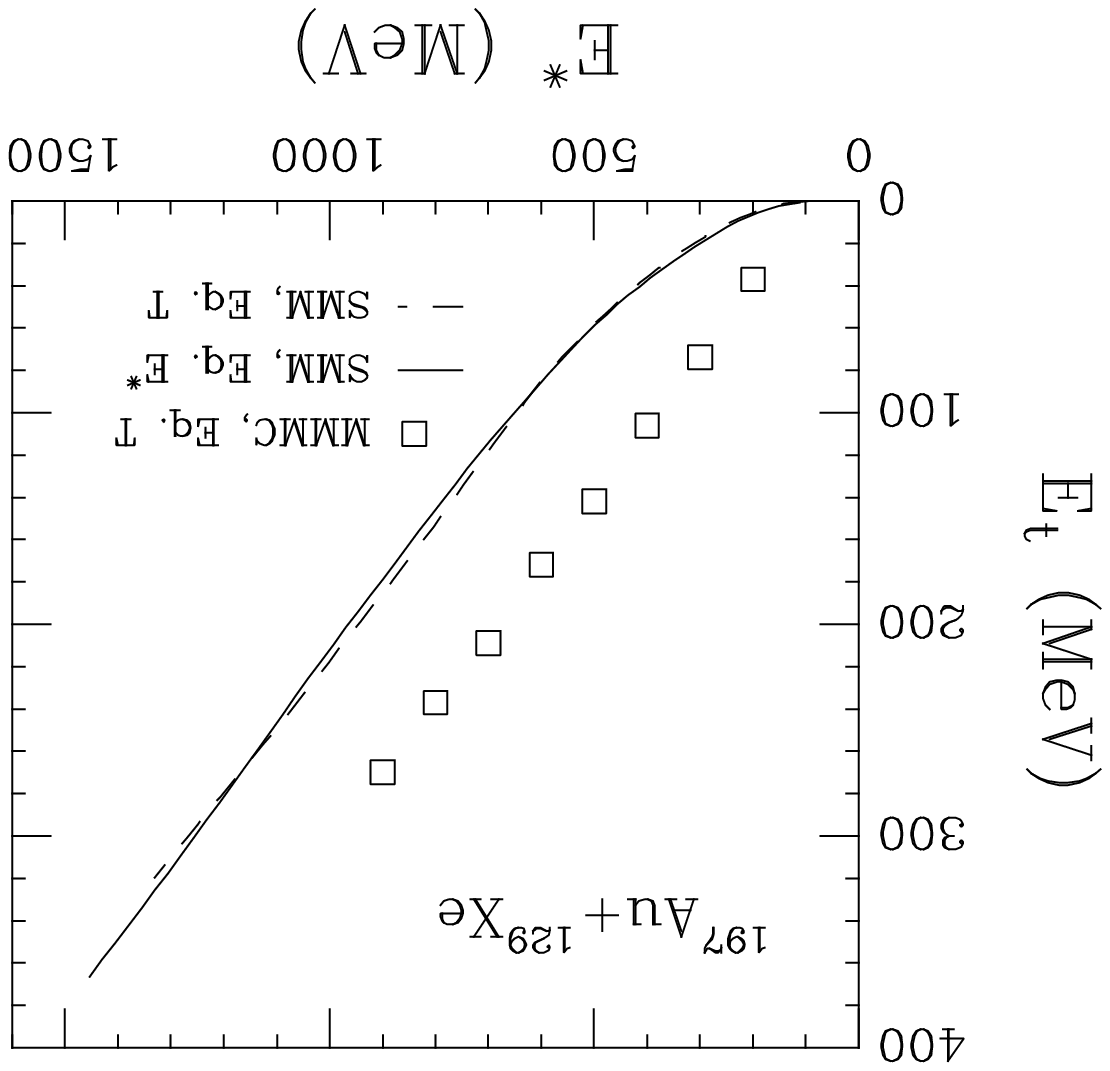}
{\footnotesize
\begin{quotation}
\noindent
Fig. 2. Dependence of the transverse energy $E_t$ on the total excitation
energy $E^*$ for the reaction $^{197}$Au+$^{129}$Xe, as predicted
by the SMM$^7$ (lines), GEMINI$^{20}$ (circles), and MMMC$^8$ 
(boxes) codes. 
The solid and dashed lines represent calculations
made assuming equal excitation energies $E^*$ or equal temperatures $T$ 
of the projectile- and target-like fragments.
\end{quotation}}
\bigskip

Fig.~2 shows the average functional dependence of $E_t$
on $E^*$, as predicted by the codes SMM\cite{smm} (solid and dashed
lines) and GEMINI\cite{gemini} (circles) for the system
$^{197}$Au+$^{129}$Xe discussed in a recent paper.\cite{mor2} 
This system is very close to the $^{209}$Bi+$^{136}$Xe system,
discussed earlier above, for which the joint multiplicity of neutrons 
and charged particles has been actually measured. Hence, it is
expected that the SMM and GEMINI models provide for an adequate description
of this system as well and, in particular, for the description
of the initial segment of the $m_n$--$m_{LCP}$ correlation ridge
parallel to the $m_n$ axis. It is the presence of this segment that
translates directly and unambiguously into the strong nonlinearity 
of the functional relationship between $E_t$ and $E^*$ and the lack
of a direct proportionality between these two quantities in the range
of excitation energies considered in the above papers.
\cite{mor1,mor2,newmor} For comparison, the predictions by the code
MMMC\cite{mmmc} are shown in Fig. 2 by squares. As expected based on
the trends exhibited by the MMMC calculations in Fig.~1, the MMMC
predicts almost a straight proportionality between $E_t$ and $E^*$.

The two almost overlapping curves in Fig.~2 represent
two extreme assumptions made regarding the excitation energy division
between Au-like and Xe-like primary fragments, equal excitation
energies or equal temperatures (excitation energies are proportional
to the masses) of projectile-like and target-like fragments.
Obviously, the results exhibit a remarkable insensitivity to the
excitation energy division between the fragments. The non-linear
functional dependence seen in Fig.~2  can be
approximated by two linear segments: $E_t=0$, for $E^*< E^*_o$, and
$E^*= a(E^*-E^*_o)$ for $E^*>E^*_o$, where the offset is $E^*_o\approx
340$ MeV and the slope is $a\approx 0.25$. 

\vspace*{10.5cm}
\includegraphics{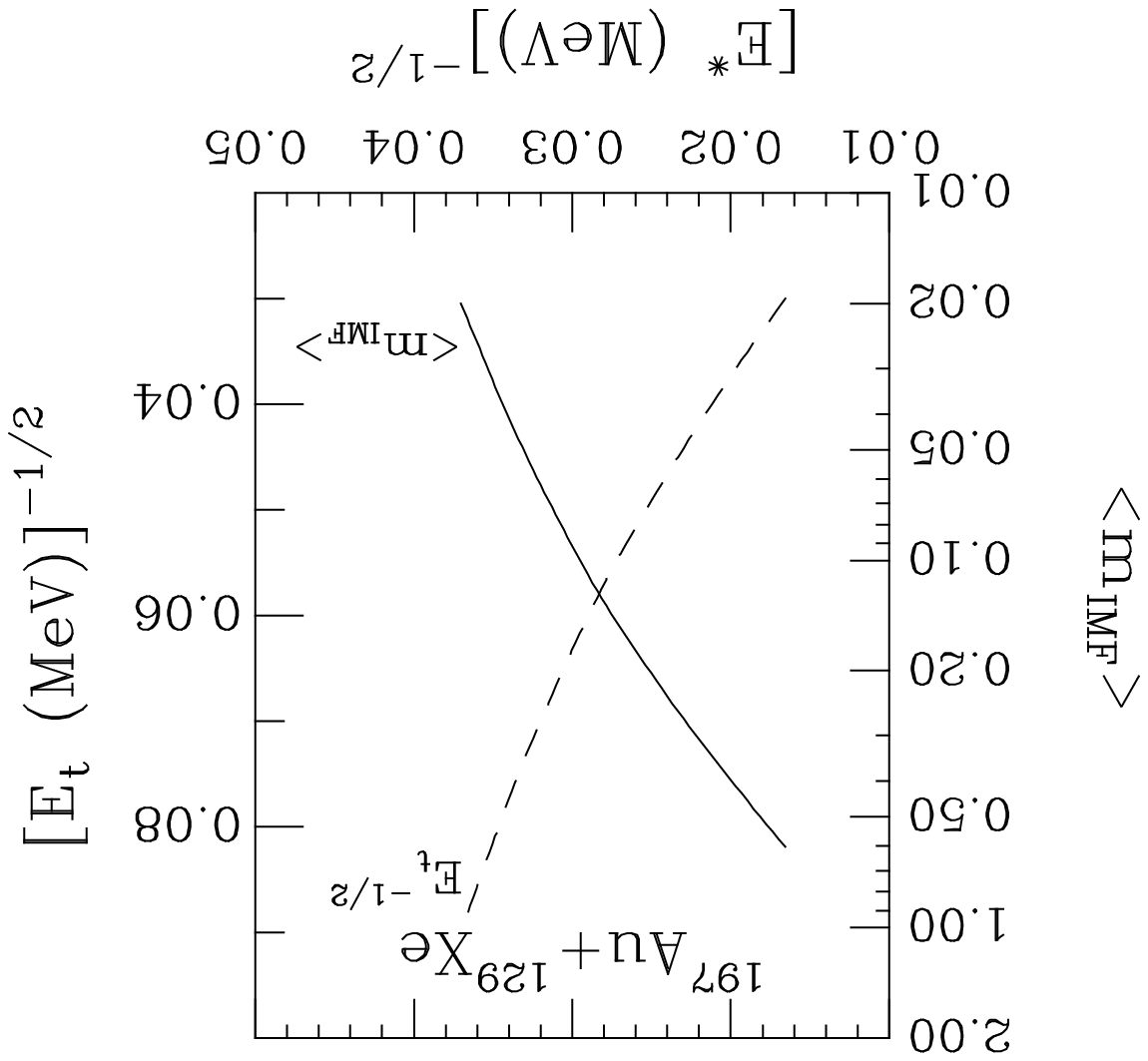}
{\footnotesize
\begin{quotation}
\noindent
Fig. 3. Nonlinearity of the
relationship between the inverse square-roots of $E_t$ and $E^*$
(dashed
line) and the resulting nonlinearity of an Arrhenius-like plot,
obtained by converting the abscissa of a previously
published$^{10,11}$
pseudo-Arrhenius plot for the system $^{197}$Au+$^{129}$Xe
(solid line).
\end{quotation}}
\bigskip

The functional dependence seen in Fig.~2 results in a
strongly nonlinear relationship between the true "thermal" Arrhenius
variable $1/T$ or its measure $1/\sqrt E^*$, on the one hand, and the
pseudo-Arrhenius variable $1/\sqrt{E_t}$, on the other hand, over the
full range of values considered in the recent
papers.\cite{mor1,mor2,botv1,botv2,newmor} This relationship is
illustrated by the dashed line in Fig.~3. Accordingly,
the persistent\cite{mor1,mor2} striking linearity  of experimental
pseudo-Arrhenius plots implies a similarly persistent nonlinearity of
such plots when converted to a true Arrhenius abscissa $1/T$ or
$1/\sqrt{E^*}$. To further demonstrate this point, the solid line in
Fig.~3 represents a typical experimental plot converted
to a representation where $1/\sqrt{E^*}$ is the abscissa variable. The
solid line is obtained based on experimental data\cite{mor2} for a
representative atomic number of $Z$ = 8 and the relationship between
$E_t$ and $E^*$ seen in Fig.~2. Obviously, the
nonlinearity of the Arrhenius-like plot in Fig.~3
contradict previous conclusions of thermal\cite{mor1,mor2} or
microcanonical\cite{botv1,botv2} scaling of IMF distributions. These
observations also demonstrate  independently that true and
pseudo-Arrhenius plots are not equivalent to each other, a fact that
has been pointed out earlier\cite{tok1,tsang,wojtfld} on somewhat
different grounds.

In conclusion, the joint distributions of neutron and charged-particle
multiplicities offer a powerful tool for scrutinizing certain
theoretical models and concepts. This has been demonstrated
here for two different examples. While the statistical-equilibrium
model  GEMINI\cite{gemini} and the Copenhagen model of simultaneous
multifragmentation SMM\cite{smm} provide a quantitative account of
these distributions, the Berlin model of microcanonical fragmentation
does not predict the characteristic
long $m_{LCP}\approx 0$ segment in
the $m_n$ {\it vs.} $m_{LCP}$ correlation ridge. The 
experimental joint $m_n$ {\it
vs.} $m_{LCP}$ multiplicity distributions confirm predictions by
phase-space models that there is an appreciable offset (in
$E^*$) in the approximately
linear functional relationship between average
transverse energy $E_t$ of charged particles and average 
total excitation 
energy $E^*$ at energies above some threshold energy of $E^*_o\approx
340$ MeV. This observation challenges conclusions regarding
thermal\cite{mor1,mor2,newmor} or microcanonical\cite{botv1,botv2}
scaling of multifragmentation, which are all based on the assumption
of a direct proportionality between $E_t$ and $E^*$.
The present study has demonstrated again the difficulty to interpret
in an unambiguous fashion isolated effects observed in complex nuclear
reactions. This experience stresses the necessity of a more holistic
analysis of energetic nuclear reactions, where many facets of
experimental observations are considered simultaneously.  Such an
approach requires the simultaneous measurement of as many experimental 
observables as possible and singling out all ``probative''
correlations that contradict some but not other theoretical or
intuitive concepts. More specifically, the present work has
demonstrated the great importance of measuring neutron data along with
those for charged reaction products.

This work was supported by the U.S. Department of Energy grant No.
DE-FG02-88ER40414.

\end{document}